\def\section{\@startsection {section}{1}{\z@}{-3.5ex plus -1ex minus
 -.2ex}{2.3ex plus .2ex}{\normalsize\bf}}
\def\subsection{\@startsection{subsection}{2}{\z@}{-3.25ex plus -1ex minus
 -.2ex}{1.5ex plus .2ex}{\normalsize\bf}}
\def\subsubsection{\@startsection{subsubsection}{3}{\z@}{-3.25ex plus
 -1ex minus -.2ex}{1.5ex plus .2ex}{\normalsize\bf}}
\def\thebibliography#1{\section*{REFERENCES\@mkboth
 {REFERENCES}{REFERENCES}}\list
 {[\arabic{enumi}]}{\settowidth\labelwidth{[#1]}\leftmargin\labelwidth
 \advance\leftmargin\labelsep
 \usecounter{enumi}}
 \def\newblock{\hskip .11em plus .33em minus .07em}
 \sloppy\clubpenalty4000\widowpenalty4000
 \sfcode`\.=1000\relax}

\def\thefigure{\@arabic\c@figure}
\def\fps@figure{tbp}
\def\ftype@figure{1}
\def\ext@figure{lof}
\def\fnum@figure{\bf Figure \thefigure}
\def\figure{\@float{figure}}
\let\endfigure\end@float
\@namedef{figure*}{\@dblfloat{figure}}
\@namedef{endfigure*}{\end@dblfloat}

\documentstyle[12pt,plenum]{article}
\begin{document}
\noindent
\bf
CHARGE-SPIN SEPARATION AND THE SPECTRAL PROPERTIES OF LUTTINGER LIQUIDS
\rm
{}~\\
{}~\\
{}~\\
{}~\\
\hspace*{1in}
\parbox{5cm}{Johannes
Voit$\footnotemark[1]$\\
{}~\\
Institut Laue-Langevin\\
38042 Grenoble (France)}\\
\footnotetext[1]{Present address: Bayreuther Institut f\"{u}r
Makromolek\"{u}lforschung (BIMF) and
Theoretische Physik~1,
Universit\"{a}t Bayreuth, D-8580 Bayreuth (Germany)}
{}~\\
{}~\\
\section*{INTRODUCTION}
There are fundamental differences between interacting
fermions in 1D and 3D. Fermi liquid theory,
describing the vicinity of the 3D Fermi surface, is based on the existence
of quasi-particles evolving out of the electron (hole) excitations
of a Fermi gas upon adiabatically switching on interactions. They are
in one-to-one correspondence with the bare particles, possess the same
quantum numbers and obey Fermi-Dirac statistics. They are robust against
small displacements away from the Fermi surface with a lifetime diverging
as $\tau \sim \mid E - E_F \mid^{-2}$. Ideally, they dominate the spectral
response with a sharp peak at $\omega = \varepsilon({\bf k})$, broadening
as $E-E_F$ increases. In addition to the quasiparticles, there
are collective (charge or spin) excitations contributing incoherent
background to the spectral function. Of course, there are borderline
cases where the quasiparticle peak is weak and most of the spectral
weight resides in the incoherent parts.

In one dimension, there are no fermionic quasiparticles in the vicinity
of the Fermi surface and the low-energy
excitations are gapless bosonic charge and
spin fluctuations\cite{solyom}. They usually propagate with different
velocities: an incoming electron decays into distinct charge and spin
excitations which spatially separate with time -- hence the name
charge-spin separation. Correlation functions decay with nonuniversal
powers as a function of $x$ and $t$ and exhibit nonuniversal singularities
as a function of $q$ and $\omega$. All these features have dramatic
consequences for the spectral properties of interacting 1D fermions which
are much less studied than those of the Fermi liquid.

Absence of fermionic quasiparticles, charge-spin separation, and nonuniversal
power law correlations are generic to 1D fermion systems but are particularly
prominent in the exactly solvable Luttinger model\cite{Lutt} whose excitations
can be viewed as a gas of noninteracting bosons. All correlation functions
can be calculated exactly. Based on case studies of Bethe Ansatz solvable
models\cite{Haldprl}, Haldane conjectured that this picture remains true,
at least in terms of renormalized bosons and up to perturbative boson-boson
interactions, for the low-energy physics of a much wider class of 1D models,
and coined the term ``Luttinger liquid'' to describe the universal low-energy
phenomenology  of gapless 1D quantum systems\cite{Haldane}. There is an
obvious analogy to the Fermi gas and Fermi liquid pictures in higher
dimensions. Haldane's conjecture has been verified extensively for many
1D lattice models by a wide variety of methods\cite{LL}.

Renewed interest for the properties of Luttinger liquids was stirred up
by Anderson's proposal that the normal state
of the high-$T_c$ superconductors could be described by a hypothetical
``tomographic'' Luttinger liquid in 2D\cite{ar}. Much of this discussion
has been based on the spectral properties of the high-$T_c$ materials
measured in photoemission\cite{olson}; a central issue is charge-spin
separation. On the other hand,
the spectral properties of even the 1D
Luttinger liquid and, in particular the signatures of charge-spin
separation there, are only poorly understood.

There is experimental evidence for Luttinger liquid behaviour
in quasi-1D organic conductors and
superconductors\cite{NMR}. In TTF-TCNQ, e.g., there are strong
$4k_F$-CDW fluctuations in addition to the usual $2k_F$ ones, indicative
of sizable Coulomb interactions. Moreover, the optical conductivity is
strongly depressed at low frequencies. Mainly based on anomalous NMR
relaxation behaviour, a strong case for a Luttinger liquid has
been made for the normal state of the organic superconductors $(TMTSF)_2X$
with $X=PF_6, ClO_4, \ldots$ (Bechgaard salts).
Most recently, photoemission studies
have been performed both on some inorganic CDW materials\cite{ne} and
$(TMTSF)_2PF_6$\cite{photem} in their normal
state. These studies generally show an intriguing absence of spectral
weight at the Fermi surface and, in angle-resolved measurements, no
dispersive feature reminiscent of quasiparticle peaks. It has been
suggested that this could be related to Luttinger liquid correlations.

In this paper, I summarize a study of the spectral function of the
Luttinger model\cite{early,Sch}.
By comparing the spinful Luttinger model with a
spinless version and with a ``one-branch'' model, a clear separation
of the influences of charge-spin separation and of the anomalous
fermion dimensions on the spectral response is possible. I also
comment on the consistency of the results presented here with the
photoemission experiments in 1D materials and its implication for their
theoretical description.

\section*{MODELS AND GREEN'S FUNCTIONS}
The Luttinger model\cite{Lutt} describes 1D right- and left-moving fermions
with linear dispersion through a Hamiltonian with the following terms:
\begin{eqnarray}
H_0 & = & \frac{1}{L} \sum_{r,k,s} v_F (rk - k_F) c^{\dag}_{rks} c_{rks} \\
 & = &  \frac{\pi v_F}{L} \sum_{\nu = \rho, \sigma} \sum_p
: \nu_+(p) \nu_+(-p) + \nu_-(-p) \nu_-(p) : \label{h0}
\end{eqnarray}
describes, equivalently, free fermions and
free charge and spin density fluctuations ($\nu = \rho,\sigma$)
with the Fermi (sound)
velocity $v_F $ on the two branches of the linear dispersion
about the two Fermi points $\pm k_F$. The
dispersion extends to infinity and all the
electronic negative energy states are filled. $: \ldots :$ denotes normal
order.
$c_{rks}$ are the fermion operators,
and the operators
for charge and spin fluctuations
\begin{equation}
\label{bos}
\nu_r(p) = \frac{1}{\sqrt{2}} \sum_{k}
(c^{\dagger}_{rk_F + k+p \uparrow} c_{rk_F + k \uparrow}
\pm c^{\dagger}_{rk_F + k+p \downarrow} c_{rk_F+  k \downarrow})
\end{equation}
obey boson commutation relations.
It is remarkable that the model can be solved exactly in presence of the
interactions.
\begin{eqnarray}
\label{h4}
H_4 & = & \frac{1}{L} \sum_{\nu = \rho, \sigma} \sum_p g_{4 \nu}(p)
: \nu_+(p) \nu_+(-p) + \nu_-(-p) \nu_-(p) : \;\;\;
\end{eqnarray}
describes
the forward scattering of fluctuations on the same branch of the spectrum;
its effect is a renormalization of the velocities $v_F \rightarrow
v_F + g_{4\nu}
/\pi$ of charge and spin fluctuations which, in general, now will differ.
\begin{eqnarray}
\label{h2}
H_2 & = & \frac{2}{L} \sum_{\nu = \rho, \sigma}
\sum_p  g_{2\nu}(p) \nu_+(p) \nu_-(-p)
\end{eqnarray}
the forward scattering between particles on different sides of the
Fermi surface, hybridizes density fluctuations on both
branches.
The effective interactions for charge and spin
\begin{equation}
g_{i\rho}  = \frac{1}{2} (g_{i\|}+g_{i\perp}) \;\;\; , \;\;\;
g_{i\sigma} = \frac{1}{2}(g_{i\|}-g_{i\perp}) \;\;\;
\end{equation}
are linear combinations of the fermions' coupling constants for parallel
and antiparallel spins.
The momentum transfer in these interactions is cut off on a scale
$1/\Lambda$.
The model is diagonalized by a Bogoliubov transformation\cite{Lutt};
the gas of noninteracting bosons emerging can be described
completely by two non-universal parameters per degree of freedom,
an exponent $K_{\nu}$,
determining the power-law decay of correlation functions,
and $v_{\nu}$, the renormalized velocities of the collective modes:
\begin{equation}
\label{ll}
K_{\nu}(p) = \sqrt{ \frac{\pi v_F + g_{4\nu}(p) - g_{2\nu}(p) }
{\pi v_F + g_{4\nu}(p) + g_{2\nu}(p) } } \;\;\;,
\hspace{0.5cm} v_{\nu}(p) = \sqrt{\left(v_F +
\frac{g_{4\nu}(p)}{\pi}\right)^2 -
\left(\frac{g_{2\nu}(p)}{\pi}\right)^2} \;\;.
\end{equation}
In all formulae below, the limit $p \rightarrow 0$ is implied in $K_{\nu}$
and $v_{\nu}$.
$K_{\sigma} = 1$ for spin-rotation invariant problems.
The retarded Green's function is
\begin{eqnarray}
G_{r,s}^{R}(x,t) & = & - i \Theta(t) < \left\{ \Psi_{rs}(x,t),
\Psi_{rs}^{\dag}(0,0) \right\}> \nonumber \\
& = & -i \frac{\Theta(t)}{2 \pi} e^{irk_Fx} \lim_{\epsilon \rightarrow 0}
\left\{ \frac{\Lambda + i (v_Ft-rx)}{\epsilon + i (v_Ft-rx)} \right.
\label{green} \\
& \times & \left. \prod_{\nu = \rho,\sigma} \frac{1}{\sqrt{\Lambda +
i(v_{\nu}t-rx)}} \left( \frac{\Lambda^2}{(\Lambda + i v_{\nu}t)^2
+ x^2}\right)^{\gamma_{\nu}} + \left(
\begin{array}{ccc}
x & \rightarrow & -x \\
t & \rightarrow & -t
\end{array}
\right) \right\}
\;\;\;,
\nonumber
\end{eqnarray}
and the spectral function is obtained as
\begin{equation}
\label{spec}
\rho_{rs}(q,\omega) = - \frac{1}{\pi} {\rm Im}G_{rs}^R (k_F+q,\omega)
\;\;\;.
\end{equation}
The first term in curly brackets originates from the momentum dependence
of the velocities $v_{\nu}$ and describes the crossover from free fermion
behaviour at very short length (time) scales to Luttinger liquid behaviour
in the asymptotic long distance (time) regime.
Two important quantities can be derived from $G(x,t)$ resp. $\rho(q,\omega)$:
the momentum distribution function
\begin{equation}
\label{nk}
n_{rs}(k) = - i \int_{-\infty}^{\infty} \: dx e^{-ikx} G_{rs}(x,0^-)
\sim (rk_F - k)^{\alpha} \;\;\; (k \approx rk_F) \;\;\;,
\end{equation}
derived from the time ordered Green's function, and the single-particle
density of states
\begin{equation}
\label{nom}
N(\omega) = \sum_{rs}
\int_{-\infty}^{\infty} \frac{dq}{2 \pi} \rho_{rs}(q,\omega)
\sim \mid \omega \mid^{\alpha} \;\;\;
(\mid \omega \mid \ll \frac{\Lambda}{v_F}) \;\;\;.
\end{equation}
Both are characterized by the effective single-particle exponent
\begin{equation}
\label{alpha}
\alpha = 2 \sum_{\nu} \gamma_{\nu}\;\;\;,  \hspace{1cm}
\gamma_{\nu}
=(K_{\nu}+K_{\nu}^{-1}-2)/8 >0 \;\;\;,
\end{equation}
measuring the
effective correlation strength (but not its sign).
For the repulsive Hubbard model $1/2 \leq K_{\rho} \leq 1$
implying $\alpha \leq1/8$\cite{LL}.

The spectral function and all results below obey to the sum rule
\begin{equation}
\label{sumrule}
\int_{-\infty}^{\infty} d \omega \rho_{rs}(q,\omega) = 1 \;\;
{\rm \; for \; all} \; q
\;\;\;,
\end{equation}
thus improving, for the
spinful Luttinger liquid, over earlier results\cite{early,Sch}.
A local sum rule is not satisfied unless $g_{4\|} = 0$ as is the
case for local interactions; in general (long range interactions), one has
\cite{early,Sch}
\begin{equation}
\label{locsum}
\pi \Lambda \int_0^{\infty} \: d \omega \left[ N(\omega) - N_0(\omega) \right]
= - \frac{g_{4\|}(p=0)}{2 \pi v_F} \;\;\;,
\end{equation}
$N_0(\omega) = 1/\pi v_F$ being the noninteracting density of states.
In the results given below, $\alpha$ is
used to measure the strength of the (repulsive) interactions. We obtain
$K_{\rho}$ and $v_{\nu}$ using Eqs.~(\ref{ll}) and (\ref{alpha})
and assume $g_{4\perp}
=g_{2\perp}=g_{2\|}$ finite while $g_{4\|}=0$.
This guarantees a local sum rule for the Luttinger model
(likely to be satisfied by a more complete theory) and is fully compatible
with spin rotation invariance.

Two important simplifications are possible:
(i) Spinless fermions, involving only charge degrees of freedom.
This limit is formally obtained by setting $g_{i\perp} = 0$.
The Green's function is obtained
from Eq.~(\ref{green}) by $(\gamma_{\sigma},v_{\sigma}) \rightarrow
(\gamma_{\rho},v_{\rho})$.
$g_2 \neq 0$ introduces a nontrivial
power-law decay of the correlation functions described by an
exponent $K_0 \neq 1$. (ii) $g_2 = 0$
but $g_{4\perp} \neq 0$ yielding $v_{\rho} \neq v_{\sigma}$ but $K_{\nu} = 1$.
This is a minimal model for charge-spin separation: the correlation
exponents $K_{\nu}$ are the same as for free fermions but the velocities
of charge and spin excitations differ. The
two branches are now independent
and we have a ``one-branch
Luttinger liquid''. For the Green's function, simply set $\gamma_{\nu} = 0$
in Eq.~(\ref{green}).
For more complicated
models possessing a Luttinger liquid fixed point, the
parameters $K_{\nu}$ and
$v_{\nu}$ can be calculated by a variety of methods\cite{LL}.
Another approximation frequently used is to neglect the momentum dependence
of the velocities of the collective modes, Eq.~(\ref{ll}). This simplifies
calculations but alters the high-energy physics. We shall partly use
this approximation below.

\section*{SPECTRAL FUNCTIONS}
We now discuss the spectral functions, Eq.~(\ref{spec}), for the three models
introduced. We only present figures for the new results on the spectral
properties of the spinful Luttinger liquid and refer the reader to the
earlier publications\cite{early} for the two toy problems.

With spinless fermions we can single out the influence of
the anomalous fermion dimensions. An approximate calculation giving correctly
the main features was published long ago by Luther and Peschel\cite{lupe}.
This is the generic structure: At $q=0$ ($k=k_F$) $\rho(0,\omega) \sim
\mid \omega \mid^{\alpha-1}$, i.e. a power-law divergence (or cusp-singularity
for $\alpha>1$) instead of the
$\delta$-function in Fermi liquid theory. Clearly,
as the 1D correlations increase from zero, spectral weight is pushed away
from the Fermi surface by the virtual particle-hole excitations generated
by $g_2$. Let us increase $q$. In a Fermi liquid, the $\delta$-function
would disperse with $q$ and broaden but essentially conserve its
shape. In the Luttinger liquid, $\rho(q,\omega)$ strongly deforms: There
is a power law singularity
$\rho(q,\omega) \sim \Theta(\omega - v_0 q)
( \omega - v_{0}  q )^{\gamma_{0}-1} $ at positive frequencies (for $q>0$)
and a weaker singularity $\sim \Theta(-\omega -v_0q)
(-\omega - v_0 q)^{\gamma_0}$ at negative frequencies. In the positive
frequency contribution -- particle creation above the Fermi surface --
spectral weight of an incoming particle is boosted to
higher energies by the particle-hole excitations on both branches. The negative
frequency contribution describes the destruction of
particles above the Fermi surface present in the ground state. As $q$
increases, the negative frequency part is exponentially suppressed and
all the spectral weight is transferred to positive frequencies.
In the approximation $v=const.$, the positive frequency part then
disperses without changing its shape. However
in the exact spectral function, based
on Eqs.~(\ref{green}) and (\ref{spec}), a second peak appears, as $q$
increases beyond $1/\Lambda$, at the free particle energy $v_F q$
relevant for the high-energy behaviour. Spectral
weight is simply transferred, like in communicating vessels, between
the Luttinger liquid and the free signal. This peak does not evolve
into a $\delta$-function as one might naively expect and its shape still
depends on $\alpha$.

For the ``one-branch'' Luttinger liquid ($g_2 =0$, charge-spin
separation only), one has finite spectral weight only
at positive frequencies (for $q>0$) between $v_{\sigma}q$ and $v_{\rho}q$
with inverse-square-root divergences at the edges\cite{Lutt,early}.
At higher momenta,
a logarithmic singularity appears at the unrenormalized energy $v_F q$ and
the behaviour at the edges changes into square-root cusps. Again, the
free particle $\delta$-function is not even recovered for $q \rightarrow
\infty$.
At $k_F$, the spectral function reduces to
$\delta(\omega)$
and the momentum distribution is a step function with
a unity jump at $k_F$,
in agreement with Luttinger's theorem\cite{lt}.
Although this seems to imply
a Fermi liquid it is clear that the
physical picture is quite different and that the notion of a quasiparticle
does not make sense since the $\delta$-function does not survive the
slightest displacement from the Fermi surface.
The incident electron decays into multiple particle--hole-like
charge and spin fluctuations which all live on the same branch
as the incoming fermion.
It is immediately apparent that $n(k)$ and, more
generally, any quantity depending on $k$ or $\omega$ alone will not exhibit
qualitatively new features due to
charge-spin separation. These can be manifest {\em only}
in quantities depending on {\em both} $q$ {\em and} $\omega$.

Let us now turn to the spectral properties of the spinful Luttinger liquid.
We limit ourselves to the spin-rotation invariant case
($\gamma_{\sigma}=0$) and use constant velocities.
The resulting (lengthy)
integral representation of the spectral function, satisfying the sum rule
(\ref{sumrule}),
can be evaluated numerically
to a high degree of accuracy. Fig. 1 displays the dispersion of
$\rho(q,\omega)$ for
small $q$ and $\alpha=0.125$, i.e. the value appropriate to
the $U=\infty$-Hubbard model.
It is apparent that the spectral function carries features both from
the spinless fermions (synonymous with ``anomalous fermion dimensions'')
and the one-branch problem (``charge-spin separation''). The picture
is, however, far more complicated than a simple addition of these two
functions, and multiple crossovers can take place between regimes where
one or the other feature is prominent.

At very small $q$, on the scale of the Figure, $\rho$ looks pretty much
like the spinless fermions' function. The splitting into two peaks, dispersing
with the charge and spin velocities $v_{\nu}$, is difficult to resolve there.
As $q$ increases, the negative frequency weight (very small anyway) is
transferred to positive frequency but, most importantly, the generic
two-peak structure of the spectral function becomes apparent. The exponent
of the singularity at $v_{\sigma} q$ is $2\gamma_{\rho}-1/2$ while it
is $\gamma_{\rho}-1/2$ at the $v_{\rho}q$-singularity and $\gamma_{\rho}$
at $-v_{\rho}q$. Since $\gamma_{\rho} = 1/16$ here, the correction to
the one-branch case is quite insignificant here and the charge-spin
separation aspect is clearly dominant at finite $q$. The weight above/below
$\pm v_{\rho}q$ originating from the anomalous dimensions is barely
visible.

Figure 2 presents the evolution with $\alpha$ of the spectral function.
As $\alpha$ increases, the various power-law divergences weaken and
finally transform into cusp-singularities. At the same time, the spectral
function becomes much less structured, and spectral weight is shifted
by the electronic correlations both to above/below $ \pm v_{\rho}q$ more
reminiscent of the spinless fermion problem. As the correlations increase,
the features originating from charge-spin separation are more and more
obscured by transfers of spectral weight over significant energies.
The negative frequency part raises to become a sizable fraction of the
weight at $\omega >0$.
It is interesting to note that, when $\alpha$ exceeds unity,
a change in sign of the singular term in $\rho$ occurs
so that the cusp at $v_{\rho}q$ effectively points upward and the one at
$v_{\sigma}q$ changes its sign from downward to upward.
This was not
detected in an earlier calculation using further approximations on
the short-range behaviour of the Green's function\cite{early}
indicating that $G(x,t)$ decays so rapidly at $\alpha>1$
that it effectively samples
much of the short-distance physics. The cusps are so small,
however, that no prominent feature is left at $v_{\sigma}q$.

Figure 3 displays the dispersion of a typical large-$\alpha$ spectral
function, here $\alpha=1.5$. For $q=0$, the cusp
in $\rho(0,\omega) \sim \mid \omega \mid^{\alpha-1}$ is upward. As $q$
becomes finite, there is again an asymmetric deformation of the spectral
function but the negative frequency part is much stronger than at small
$\alpha$. All cusps have
turned upward once $\alpha$ has exceeded unity. Dominant now are those
of the charge fluctuations at $\pm v_{\rho}q$ while the signal at $v_{\sigma}
q$ is barely visible. Moreover $v_{\sigma}$ is so small that the onset
of spectral weight is nearly pinned to zero frequency despite the increase
in $q$. The general appearance at large $\alpha$
is more similar to a spinless fermion spectrum. When accounting for
the dispersion in $v_{\nu}(p)$ in (\ref{green}), Figs.~1--3
will correctly describe the spectral functions for $q < 1/\Lambda$; above,
the spectral weight will be transferred into a peak at $v_Fq$.

Combining these results with information on allowed values of $\alpha$
for different models, one would conclude that local interactions
(Hubbard model) even of infinite magnitude produce surprisingly weak
correlations in the single-particle properties. The dominant deviations
from Fermi liquid behaviour then originate in charge-spin separation.
Only a finite range interaction can produce \em really strong correlations,
\rm necessarily at the expense of spectral features due to charge-spin
separation. A Coulomb $1/q^2$-potential is always in the strong correlation
limit no matter its coupling constant. For realistic
(lattice) models, deviations from the Luttinger liquid behaviour described
here will occur at finite $q$ and $\omega$; in particular, we expect that
charge and spin will not separate indefinitely at higher momenta. Finally,
the vanishing spectral weight for certain frequency ranges in Figs.~1-3
is a direct consequence of 1D kinematics (similarly to the forbidden
low-energy region in the particle-hole excitation spectrum for $0 \leq
\mid q \mid leq 2k_F$)
as discussed elsewhere\cite{early};
it is thus robust against finite temperature.

\section*{PHOTOEMISSION OF QUASI-1D CONDUCTORS}
These results are relevant for photoemission experiments\cite{ne,photem}
on quasi-1D systems believed to be Luttinger liquids\cite{NMR}.
The prototype 1D ``metal'' $(TMTSF)_2PF_6$
undergoes a spin density wave or superconducting transition at low
temperature, depending on pressure. There is evidence for important repulsive
interactions, and NMR in the metallic state has been analyzed in terms
of Luttinger liquid correlations.
A high-resolution photoemission experiment\cite{photem}
with excellent energy and moderate angular resolution failed
to detect, in the normal state,
any signature of a Fermi edge or any feature dispersing as the
momentum transfer $q$ is varied. Only a broad peak is observed at about
$-1eV$ at the bottom or even outside the conduction band of $(TMTSF)_2PF_6$.
Surprisingly similar behaviour is observed in most of the
quasi-1D CDW materials\cite{ne} while a Fermi edge is clearly detected in the
more 2D system $1T-TaS_2$.

To what extent is the spectral response of $(TMTSF)_2PF_6$ consistent with
a Luttinger liquid hypothesis? Due to the bad angular resolution, one would
expect a frequency signal reminiscent of the density of states ($\mid \omega
\mid^{\alpha}$) with dispersion characteristic of that of the divergences
in $\rho(q,\omega)$.
It is manifestly inconsistent if one considers
small values of $\alpha$ e.g.~appropriate for the Hubbard model. Here theory
would predict both a pronounced Fermi ``edge'' as well as dispersion of the
signal(s). None is observed. The experiments are more consistent, however,
with a large-$\alpha$ (in excess of unity) Luttinger liquid, indicative of
strong long-range interactions. In this case, all of the spectral weight
at the Fermi surface is pushed to higher energies, and the density of states
starts off flat. Moreover, as suggests Fig.~3 and has been discussed
elsewhere\cite{early}, no significant dispersion is expected
so long as $\mid q \Lambda \mid \leq 1$. The key feature
used in this argument is the observed absence of spectral weight close to
the Fermi surface. The maximum observed at $-1eV$ may well be due to either
higher-energy excitations of a more complete theory (the Luttinger liquid
picture is likely irrelevant on that energy scale) or molecular orbitals
of the parent molecules below the conduction band of the solid, and need
not be related to maxima implied by local sum rules, compensating the depletion
of spectral weight at small $\omega$.

The classification of the Bechgaard salts as large-$\alpha$ Luttinger liquids
is consistent with other experiments. One can go back from $\alpha$ to
$K_{\rho}$ via (\ref{alpha}), assuming repulsive interactions, and then
derive the exponents of \em all \rm other correlation functions. Recent
NMR experiments give $K_{\rho} \sim 0.15$\cite{NMR}
and imply $\alpha=1.25$ consistent
with photoemission. Another prediction based on such large
$\alpha$ is that the $4k_F$-CDW response should be more divergent
than the one of $2k_F$-CDWs and SDWs. $4k_F$-CDWs
have not been observed to date for $(TMTSF)_2PF_6$ -- but neither in the
related $(TMTTF)_2X$-series where $4k_F$-localization is firmly
established. With these caveats in mind, one would characterize
the Bechgaard salts as large-$\alpha$ Luttinger liquids, close to
a charge localization transition (Wigner crystallization) but still on the
metallic side; evidence
for important charge localization effects has also been produced with infrared
spectroscopy\cite{pecile}.
$(TMTTF)_2X$ would then be really on the edge
since it undergoes localization at low temperatures.

Most surprising is the apparent universality of spectral response
in the normal state of quasi-1D
systems. The inorganic CDW-materials are believed to be dominated by the
(attractive) electron-phonon interaction. It has
been suggested\cite{ne} that Luttinger liquid correlations could be the
origin of the absence of low-energy spectral weight in the CDW-materials.
While the exponent $\alpha$ is not sensitive to the sign of the interactions,
a Luttinger model with strongly attractive forward
scattering would either show
superconducting fluctuations or phase separation quite to the contrary of
the observations. CDWs are a consequence of relevant
electron-phonon backscattering,
however, and then model calculations\cite{phon} show that a spin
gap $\Delta_{\sigma}$
is opened on the 1D chain. In such a situation, there is no spectral
weight for
$\mid \omega \mid < \Delta_{\sigma}$. Moreover, so long as the charge
degrees of freedom remain massless, they will wipe out any spectral signature
of the $(\omega - \Delta_{\sigma})^{-1/2}$-divergence in
the density of states of the
spin fluctuations, and the onset of $N(\omega)$ at $\Delta_{\sigma}$ will
be smooth\cite{early}. The physics is best seen
in the $U<0$-Hubbard model where a similar situation occurs: electrons
form bound local singlet pairs, and such a pair must be broken before
a single electron can be photoemitted. The minimal energy to pay,
in this ``spin-paired Luttinger liquid'' is the
binding energy $\Delta_{\sigma}$.

\section*{SUMMARY}
We have discussed the spectral functions of the 1D Luttinger model and
paid care to separate the influences of the anomalous dimensions of the
fermion operators and of charge-spin separation. Charge-spin separation
can only be measured in the full dynamical ($q$- and $\omega$-dependent)
spectral functions, and splits the spectral response into two peaks dispersing
with the energies of the charge and spin fluctuations. The anomalous
correlations transfer spectral weight to frequencies above/below $\pm
v_{\rho}q$. At small correlation strength $\alpha$, charge-spin separation
is the dominant effect; when $\alpha$ increases, its spectral signatures
are gradually wiped out by transfers of spectral weight over significant
frequency scales. Comparing to photoemission data on the quasi-1D
organic superconductor $(TMTSF)_2PF_6$, consistency would require
to place this material in the large-$\alpha$ ($>1$) limit and describe
it as an ``almost localized Luttinger liquid''. This assignment is
consistent with information from NMR and infrared spectroscopy.

\noindent
{\bf Figure Captions} \\
{\bf Figure 1:} Luttinger liquid spectral function $\rho_+(q,\omega)$
for $q \geq 0$ and $\alpha=0.125$. The $\omega < 0$-part has been
multiplied by $10$ for clarity. \\
{\bf Figure 2:} Evolution with $\alpha$ of the spectral function
$\rho_+(q=0.5,\omega)$.\\
{\bf Figure 3:} Dispersion of the spectral function $\rho_+(q,\omega)$ for
$\alpha=1.5$.

\end{document}